\documentclass[manuscript]{aastex}
\usepackage{amsfonts}
\usepackage{mathrsfs}
\usepackage{natbib}
\usepackage{booktabs}
\usepackage{threeparttable}

\newcommand{\fexxi}{Fe \scriptsize{XXI} \normalsize}
\newcommand{\feii}{Fe \scriptsize{II} \normalsize}
\newcommand{\fexii}{Fe \scriptsize{XII} \normalsize}
\newcommand{\fexix}{Fe \scriptsize{XIX} \normalsize}
\newcommand{\ci}{C \scriptsize{I} \normalsize}
\newcommand{\oi}{O \scriptsize{I} \normalsize}
\newcommand{\siii}{Si \scriptsize{II} \normalsize}
\newcommand{\mgii}{Mg \scriptsize{II} \normalsize}

\begin{document}

\title{Observational Evidences of Electron-driven Evaporation in two Solar Flares}

\author{D. Li\altaffilmark{1,2}, Z. J. Ning\altaffilmark{1}, and
        Q. M. Zhang\altaffilmark{1}}
\affil{$^1$Key Laboratory of Dark Matter and Space Astronomy, Purple
Mountain Observatory, CAS, Nanjing 210008, China}
\affil{$^2$University of Chinese Academy of Sciences, Beijing
100049, China} \altaffiltext{3}{Correspondence should be sent to:
lidong@pmo.ac.cn.}

\begin{abstract}
We have explored the relationship between hard X-ray (HXR) emissions
and Doppler velocities caused by the chromospheric evaporation in
two X1.6 class solar flares on 2014 September 10 and October 22,
respectively. Both events display double ribbons and {\it Interface
Region Imaging Spectrograph} ({\it IRIS}) slit is fixed on one of
their ribbons from the flare onset. The explosive evaporations are
detected in these two flares. The coronal line of \fexxi 1354.09
{\AA} shows blue shifts, but chromospheric line of \ci 1354.29 {\AA}
shows red shifts during the impulsive phase. The chromospheric
evaporation tends to appear at the front of flare ribbon. Both
\fexxi and \ci display their Doppler velocities with a
`increase-peak-decrease' pattern which is well related to the
`rising-maximum-decay' phase of HXR emissions. Such anti-correlation
between HXR emissions and \fexxi Doppler shifts, and correlation
with \ci Doppler shifts indicate the electron-driven evaporation in
these two flares.
\end{abstract}


\section{Introduction}
Solar flares are drastic explosive phenomena on the Sun. They are
able to release a huge amount of energy ($\sim$10$^{32}$ erg) in a
typical time scale of tens of minutes. Based on the standard model,
the flare energy is transferred from the magnetic energy. Magnetic
reconnection is thought to be the primary energy release mechanism
that heats the plasmas and accelerates the bi-directional nonthermal
electrons in the solar atmosphere. This is known as the CSHKP model
\citep{Carmichael64,Sturrock66,Hirayama74,Kopp76}. These nonthermal
electrons which guided by the reconnected magnetic field lines, not
only travel to the interplanetary space but also precipitate into
the lower corona and upper chromosphere, where they lose energy to
produce radiation through Coulomb collisions with the denser medium.
This has been known as the `thick-target' model for the hard X-ray
(HXR) emission \citep{Brown71,Syrovat72}. Observations show that
only a small fraction of the energy is lost through
extreme-ultraviolet (EUV) radiation
\citep{Emslie78,Milligan14,Milligan15}. The bulk of the energy heats
the local chromospheric material rapidly up to a typical temperature
of $\sim$10 MK. Then the resulting overpressure can drive the mass
flow upward along the loop at speeds of a few hundreds kilometers
per second. The hot plasmas fill the flaring loops in a process
called `chromospheric evaporation'
\citep{Brown73,Acton82,Fisher85a,Fisher85b,Liu06,Ning09,Ning10,Zhang13,Milligan15},
resulting into soft X-ray (SXR) emission rising up. Substantial
evidences of chromospheric evaporation have been reported in X-ray
\citep[e.g.,][]{Liu06,Ning09,Ning10,Ning11a,Ning11b,Nitta12,Zhang13},
EUV
\citep[e.g.,][]{Doschek80,Feldman80,Antonucci82,Ding96,Teriaca06,Milligan06a,Milligan06b,Milligan09,Veronig10,Chen10,Li11,Doschek13,Brosius13}
and radio \citep{Aschwanden95,Karlicky98,Ning09} emissions. Previous
observations reveal that HXR emission tends to rise with the double
footpoint sources along the flaring loop legs, eventually merging
them into a single source at the same position of the loop-top
source \citep{Liu06,Ji07,Ji08,Ning09,Ning10,Ning11a,Ning11b}. This
is because the dense materials from the chromosphere rise upward
along the loops, following the movement of the HXR emission targets.
The flare shows the mergence speed of around 200 km s$^{-1}$. Such
large value of the mass evaporation is clearly demonstrated by the
blue shifts at the EUV spectral observations of the coronal lines
\citep{Feldman80,Ding96,Berlicki05,Teriaca06,Brosius07,Milligan09,Veronig10,Chen10,Li11,Doschek13,Brosius13,Tian14,Young15}.
Spectral images exhibit that the blue shifts tend to appear at the
outsides of the flare ribbons \citep{Czaykowska99,Li04}. Evaporation
materials with high temperature rise upward to disturb the coronal
plasma, which results into the radio emission suddenly suppressed on
the radio dynamic spectra, especially at the decimeter range. A
high-frequency cutoff drifting to lower frequency is thought to be
the signature of chromospheric evaporation
\citep{Aschwanden95,Karlicky98,Ning09}

From the observations, there are two types of chromospheric
evaporation. Evaporation is said to proceed `gently' when the
chromosphere plasmas lose energy via a combination of radiation and
low-velocity hydrodynamic expansion. Emission lines formed at
temperature characteristic of the atmosphere from chromosphere
through transition region to corona all appear blue shifted
\citep{Milligan06b,Brosius09,Raftery09,Li11}. Evaporation is
regarded to proceed `explosively' when the chromosphere is unable to
radiate energy at a sufficient rate and consequently expands at high
velocities into the overlying flare loops. The overpressure of
evaporated material also drives low-velocity downward motion into
the underlying chromosphere, in a process known as `chromospheric
condensation' \citep{Wulser94,Czaykowska99,Kamio05,Del06,Teriaca06}.
In this case, emission lines formed at temperature characteristic of
the upper chromosphere and transition region all appear red shifted,
while hotter lines from the corona appear blue shifted
\citep{Fisher85a,Fisher85b,Milligan06a,Del06,Brosius09,Raftery09,Li11}.
Spectral observations show the red shifted velocity of $\sim$20 to
40 km s$^{-1}$, which is an order smaller than the blue shifted
value ($\sim$200 km s$^{-1}$). This is because the plasma density of
the underlying lower chromosphere is much higher than that of the
overlying corona.

Up to now, there are two viewpoints about how to drive the
evaporation in the  literatures. One is the electron-driven, while
another is the thermal conduction driven. The former emphases that
the non-thermal energy of nonthermal electrons play an important
role in the evaporation
\citep{Fisher85a,Fisher85b,Milligan09,Tian14}, while the latter
focuses on the thermal energy directly driven
\citep{Fisher85a,Falewicz09}. In this paper, using the observations
from {\it Fermi} Gamma-ray Burst Monitor (GBM), {\it Reuven Ramaty
High Energy Solar Spectroscopic Imager} ({\it RHESSI}), Atmospheric
Imaging Assembly (AIA) aboard {\it Solar Dynamics Observatory} ({\it
SDO}) and {\it Interface Region Imaging Spectrograph} ({\it IRIS}),
we explore the relationship between HXR emissions and Doppler
velocities caused by evaporation during the impulsive phase in two
solar flares in order to detect the observation evidences of the
electron-driven chromospheric evaporation.

\section{Observations and Data Analysis}

\subsection{Observations}
Two X1.6 solar flares are selected to study in this paper. Firstly
they are well covered by the {\it IRIS} \citep{Dep14} spectral
observations. {\it IRIS} slit is fixed on the flare ribbon during
the impulsive phase, which gives us an opportunity to detect the
whole history of Doppler velocities caused by evaporation. Secondly,
HXR emissions are also well observed by {\it Fermi} \citep{Meegan09}
or {\it RHESSI} \citep{Lin02}. One event takes place in NOAA AR
12158 on 2014 September 10. It starts at 17:21 UT and reaches its
maximum at 17:45 UT on the {\it GOES} SXR light curves. Another
occurs in NOAA AR 12192 on 2014 October 22. It starts at 14:02 UT
and peaks at 14:28 UT on the SXR emissions.

\begin{figure}
\epsscale{1.0} \plotone{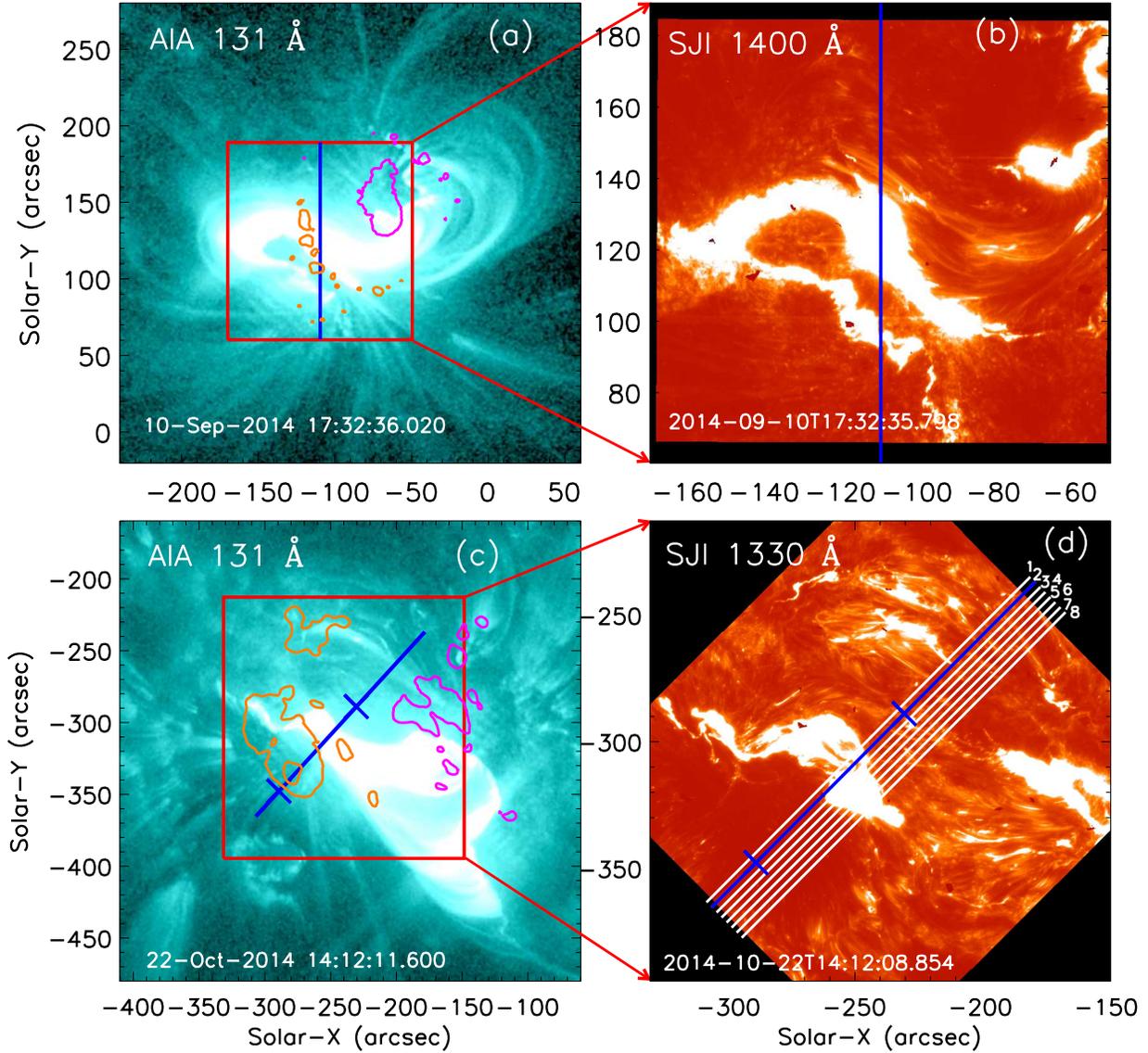} \caption{Top: AIA 131 {\AA} (a) and
SJ 1400 {\AA} (b) images at 17:32 UT on 2014 September 10. Bottom:
AIA 131 {\AA} (c) and SJ 1330 {\AA} (d) images at 14:12 UT on 2014
October 22. The blue and white lines represent the {\it IRIS} slit
positions. The contours indicate the magnetic fields from HMI at the
levels of 800 (purple) and -800 (orange) G. The red boxes mark the
{\it IRIS} SJI regions. Two short blue lines on the {\it IRIS} slit
in panels (c, d) mark the flare ribbon shown in Fig.~\ref{vel2}.
\label{image}}
\end{figure}

Fig.~\ref{image} shows the {\it SDO}/AIA \citep{Lemen12} 131 {\AA}
images (a, c) and {\it IRIS}/SJ images (b, d) of these two flares.
The contours on the AIA 131 {\AA} images represent the line-of-sight
magnetic fields from Helioseismic and Magnetic Imager (HMI)
\citep{Schou12} aboard {\it SDO}. The levels are set at 800 (purple)
and -800 (orange) G, respectively. {\it IRIS} slit is fixed along
the solar North-South direction on one ribbon of 2014 September 10
flare. The cadence is 9.4 s. For the 2014 October 22 flare, {\it
IRIS} slit is along the 45 degree to the North-South direction. {\it
IRIS} detects the flare ribbon in the `raster' mode. Each raster has
eight steps (marked by the number on the SJI 1330 {\AA} image). Each
step has a cadence of 16.4 s and a distance of $\sim$2\arcsec. Thus
each raster has a duration of $\sim$131 s.

Both events display the double ribbons at SJI 1400 {\AA} or 1330
{\AA} images, as shown in Fig.~\ref{image} (b, d). The 2014
September 10 flare shows one short ribbon around the positive filed
region, while another is long with a curved shape around the
negative filed region. This long ribbon is dynamic and propagating
toward the South-East direction, subsequently crosses the {\it IRIS}
slit during the flare impulsive phase. Fig.~\ref{ribbon} gives the
time evolution of this ribbon on AIA 1600 {\AA} images. The arrow
marks the ribbon propagation direction. This ribbon also exhibits
strong Quasi-Periodic Pulsations \citep[i.e.,][]{Liting15,Li15a}.
Similar to the 2014 September 10 flare, one ribbon of the 2014
October 22 flare is around the positive filed region, and the other
ribbon at the negative filed region is detected by the {\it IRIS}
slit during the flare impulsive phase.

\subsection{Spectral fitting}
Fig.~\ref{fit1} shows the {\it IRIS} spectral profiles of three
windows at `1343' (a), `\fexii' (b), and `\oi' (c) for the 2014
September 10 flare. {\it IRIS} has a spectral scale of $\sim$25.6
m{\AA}/pixel for these three windows. The spectral data has been
firstly calibrated and processed with the routines of
iris\_orbitvar\_corr\_l2.pro and iris\_prep\_despike.pro in the
solar software (SSW) package. The first routine is used to correct
{\it IRIS} spectral image deformation caused by the spacecraft
orbital variation \citep{Tian14b,Cheng15}. The second one is a
generalized despiking tool for {\it IRIS} data. It could identify
and remove the bad pixels through the iterative approach. At the
flare onset (i.e., 17:28:43 UT), the spectral window at `\oi' is
characterized by many narrow, bright emission lines from neutral and
singly ionized species, as well as molecular fluorescence lines.
These emission lines blend with the broad line of \fexxi 1354.09
{\AA} \citep[seen also.,][]{Li15a,Polito15,Tian14,Tian15,Young15},
which is a typical coronal line to be used to detect the
chromospheric evaporation. However, these blending chromospheric
emission lines must be extracted before determining the \fexxi
intensity. According to the characteristics of {\it IRIS} spectral
data, three steps are followed.

Firstly, to determine the line centers and widths (FWHM: the full
width at half maximum) at `\oi' window, including blending lines and
\fexxi. There are seven blending lines of \fexxi, including \feii
1353.07, 1354.06 {\AA}, \siii 1352.69, 1353.78 {\AA}, \ci 1354.29
{\AA}, and two unidentified lines at 1353.40, 1353.61 {\AA}, such as
marked by the red vertical ticks (except for \ci) in Fig.~\ref{fit1}
(c). These lines are bright in the active regions while \fexxi is
quiet. Therefore, their line centers and widths can be detected from
the single-Gaussian fitting in the active regions. During the solar
flare, their centers and widths are constrained in a range to do
spectral fitting based on the previous observations
\citep[e.g.,][]{Curdt01,Curdt04}. Taking the emission line of \siii
1352.69 {\AA} for example, its center and width during the flare are
same as the values from the active regions but constrained, such as
line center at 1352.69$\pm$0.102 {\AA} (ranging from 1352.588 to
1352.792 {\AA}), and the maximum width of 260 m{\AA}, as listed in
table~\ref{table}. \ci has the strongest emission among these seven
lines. Previous studies
\citep{Doschek75,Cheng79,Mason86,Innes03a,Innes03b} have identified
its rest wavelength at 1354.29 {\AA} and its narrow width, which is
assumed with maximum value of 130 m{\AA} for the flare spectral
fitting in this paper. Its line center is set at the range of
1354.29$\pm$0.26 {\AA} for the flare spectral fitting. Based on the
fact that \fexxi is a broad line
\citep[see.,][]{Doschek75,Cheng79,Mason86,Innes03a,Innes03b,Tian14,Tian15,Li15a}.
Its line center is set as 1354.29$\pm$1.28 {\AA}, which almost cover
the whole `\oi' window, and its line width is assumed with a minimum
value of 230 m{\AA}, as listed in Table~\ref{table}.

Secondly, to tie the blending line intensities from other similar
isolated lines during the flare. Fig.~\ref{fit1} (c) shows coronal
line of \fexxi 1354.29 {\AA} blending with these seven emission
lines for the 2014 September 10 flare. There is no way to determine
their flare intensities only at `\oi' window. However, {\it IRIS}
has the spectra at other windows, i.e., `1343', `\fexii'. They also
have the emission lines which behave similarly to the blending lines
at `\oi' window, such as H$_2$ 1342.83 {\AA} at `1343' window, \siii
1350.13 {\AA} at `\fexxi' window, and \feii 1354.80 {\AA} at `\oi'
window. Their intensities can be used to tie the intensities of the
six blending lines at `\oi' window during the flare eruption, as
listed in Table~\ref{table}. This is because \siii 1352.69, 1353.78
{\AA} at `\oi' window have a similar behavior as \siii 1350.13 {\AA}
at `\fexii' window, and \feii 1353.07, 1354.06 {\AA} at `\oi' window
have a similar behavior as \feii 1354.80 {\AA} at `\oi' window, and
Unknown 1353.40, 1353.61 {\AA} have a similar behavior as H$_2$
1342.83 {\AA} at `1343' window during the flare. These three
emission lines are isolated and their intensities can be determined
by a single-Gaussian fitting

Thirdly, to determine the line parameters of \fexxi and \ci during
the flare using the multi-Gaussian fitting. Fig.~\ref{fit1} gives
the six blending lines (except for \ci) and three isolated emission
lines marked with the red vertical ticks at three {\it IRIS}
spectral windows. \fexxi and \ci are shown by the turquoise and
magenta profiles. There are also another four isolated and bright
lines marked with blue vertical ticks, such as Unknown 1348.34,
1348.60, and 1349.65 {\AA} at `\fexii' window, and \feii 1354.91
{\AA} at `\oi' window. In total, these 15 lines (i.e., H$_2$ 1342.83
{\AA}, Unknown 1348.34 {\AA}, Unknown 1348.60 {\AA}, Unknown 1349.65
{\AA}, \siii 1350.13 {\AA}, \siii 1352.69 {\AA}, \feii 1353.07
{\AA}, Unknown 1353.40 {\AA}, Unknown 1353.61 {\AA}, \siii 1353.78
{\AA}, \feii 1354.06 {\AA}, \fexxi 1354.09 {\AA}, \ci 1354.29 {\AA},
\feii 1354.80 {\AA}, \feii 1354.91 {\AA}) superimposed on a linear
background are used to do the multi-Gaussian fitting at three {\it
IRIS} spectral windows simultaneously, such as `1343', `\fexii' and
`\oi' windows. In our fitting method, both \fexxi and \ci have free
intensities, almost free line centers and widths, as listed in
Table~\ref{table}. The six blending lines of \fexxi have constrained
positions and widths and tied intensities. The other seven isolated
lines have constrained positions and widths, but free intensities.
Fig.~\ref{fit1} shows one example of such multi-Gaussian fitting.
The black profiles are the observational spectra at the positions of
about 64.7\arcsec\ (marked by short orange line) on the slit for the
2014 September 10 flare. The brown profiles are 15 fitting lines and
the green line is the background. In this case, the intensities,
centers and widths of 15 lines can be measured by the multi-Gaussian
fitting at any time and at any positions along the {\it IRIS} slit.
Fig.~\ref{fit1} also shows that there are still some other unknown
emission lines at these three spectral windows, such as 1342.09,
1344.08, 1348.03, 1350.75, 1352.02, 1355.64 {\AA}. They are located
at the edges of the spectral windows and do not affect the spectral
fitting to detect \fexxi and \ci. Therefore they are not used to
spectral fitting in this paper.

Fig.~\ref{fit2} gives four examples of the multi-Gaussian fitting
for the flares on 2014 September 10 (top) and October 22 (bottom),
respectively. The short vertical lines represent the rest
wavelengths of \fexxi (turquoise) and \ci (magenta). It is well
known that \fexxi is a hot coronal line with a formation temperature
of about 11 MK (log T $\approx$ 7.05), which results into \fexxi
absent on the quiet Sun. Therefore, the rest wavelength of \fexxi is
not determined from the quiet regions. Recent studies from {\it
IRIS} observations show that \fexxi has a rest wavelength range
between 1354.08 and 1354.10 {\AA}
\citep[e.g.,][]{Tian14,Tian15,Young15,Graham15,Polito15}. In this
paper, we use their average value of 1354.09 {\AA} as \fexxi rest
wavelength. \ci is a typical chromospheric line with a formation
temperature of around 10$^4$ K (log T $\approx$ 4.0)
\citep{Huang14}, and its rest wavelength can be determined from the
emissions at the quiet regions, as the dashed profile shown in
Fig.~\ref{fit2} (b), which plots the spectral profile average with
10 pixels around the position marked by the short black line. The
spectra between 1353.66 {\AA} and 1354.68 {\AA} around \ci are
shown. In this paper, \ci has a rest wavelength around 1354.29
{\AA}.

\subsection{Fitting parameters}
Using the method mentioned above, the intensities, line centers and
widths of \fexxi and \ci can be determined from the multi-Gaussian
fitting along the {\it IRIS} slit. Fig.~\ref{vel1} shows the
space-time diagrams of the intensities (a, c) and Doppler velocities
(b, d) of \fexxi and \ci from 17:12 to 17:58 UT for the 2014
September 10 flare. There are strong EUV line emissions at about
60\arcsec\ $-$ 80\arcsec\ and 35\arcsec\ $-$ 40\arcsec\ along the
slit. These two regions correspond to two propagation fronts (around
120 arcsec and 100 arcsec along the slit at 17:26 UT) of the curved
ribbon in Fig.~\ref{ribbon} (b), and the north region is the flare
ribbon marked by the arrow. The spectral profiles at two positions
of 64.7\arcsec\ and 60.6\arcsec\ on the slit but three different
time are given in Figs.~\ref{fit1} and~\ref{fit2} (a, b). The 2014
September 10 flare ribbon starts to cross {\it IRIS} slit at about
17:25 UT although there are weak emissions from 17:12 UT. The flare
ribbon front displays a narrow point at the beginning, then expands
rapidly to a wide range of $\sim$25\arcsec\ after 17:40 UT along the
slit. As mentioned before, this process suggests the flare ribbon
propagation across the {\it IRIS} slit.

Doppler velocity is detected by the fitting line center separation
from the rest wavelength, whatever \fexxi or \ci. Fig.~\ref{vel1}
(b) shows that the flare ribbon starts blue shifts, then red shifts
at coronal line of \fexxi, which is due to the chromospheric
evaporation at the beginning of the flare, then the hot materials
fall back to the chromosphere along the flare loop after cooling.
This is consistent with the standard flare model and previous
findings \citep{Czaykowska99,Li04}, and the evaporation appears at
the outer edge of the flare ribbon. On the other hand, the
evaporation tends to appear at the front of flare ribbon. The
evaporation speed can reach $\sim$230 km s$^{-1}$, while the falling
speed is $\sim$25 km s$^{-1}$ in the corona. Fig.~\ref{vel1} (b)
shows that the evaporation roughly has a time scale of more than 10
minutes. Fig.~\ref{vel1} (d) gives the space-time diagram of \ci
Doppler velocities. Different from coronal line of \fexxi, the
chromospheric line of \ci exhibits red shifts all during the flare,
indicating that the 2014 September 10 flare is the explosive
evaporation. There is a velocity peak of \ci at the same time as
\fexxi blue shift. The velocity can reach $\sim$27 km s$^{-1}$.
After blue shift, \fexxi displays the similar value of red shift as
\ci.

Same as Fig.~\ref{vel1}, Fig.~\ref{vel2} gives the space-time
diagrams along the {\it IRIS} slit of the intensities and Doppler
velocities of \fexxi and \ci for the 2014 October 22 flare. As noted
earlier, this event is observed in `raster' mode with 8 steps. Thus,
we can get 8 space-time diagrams at 8 step positions, respectively.
Fig.~\ref{vel2} just shows the space-time diagram at the second
step. In this case, the slit cadence is $\sim$131 (16.4$\times$8) s.
The spectral profiles at two positions of about 59.2\arcsec\ and
47.4\arcsec\ on the slit are shown in Fig.~\ref{fit2} (bottom).
Similar to the 2014 September 10 flare, this event is the explosive
evaporation. The coronal line of \fexxi displays blue shift at the
ribbon onset, then red shift, while the chromospheric line of \ci
exhibits red shift during the whole flare. The evaporation speed can
reach $\sim$145 km s$^{-1}$, and the falling speed is around 20 km
s$^{-1}$. The evaporation timescale is roughly estimated more than
10 minutes.

\section{Results}
In order to study the relationship between the HXR emission and the
evaporation speed, Fig.~\ref{light} plots the X-ray light curves and
the time evolution of Doppler velocities at \fexxi and \ci for both
events. The 2014 September 10 flare is well detected by {\it Fermi},
but missed by {\it RHESSI}, which detects the 2014 October 22 flare.
Fig.~\ref{light} (a) shows the {\it GOES} 1.0$-$8.0 {\AA} flux
(black dashed lines) and {\it Fermi}/GBM light curves at 5 energy
channels, such as 4.6$-$12.0 keV, 12.0$-$27.3 keV, 27.3$-$50.9 keV,
50.9$-$102.3 keV, and 102.3$-$296.4 keV. They are detected by the n2
detector, whose direction angle to the Sun is stable
($\sim$60$^{\circ}$) before 17:45 UT, then its angle changes to the
Sun to produce an X-ray peak, which is not real. There is a data gap
after 17:54 UT. The time resolution of {\it Fermi} is 0.256 s, but
becomes a higher value (0.064 s) automatically in the flare state.
We rebin all the data into an uniform resolution of 0.256 s here, as
shown in Fig.~\ref{light} (a). Fig.~\ref{light} (c, e) gives the
time evolution of \fexxi and \ci Doppler velocities at two positions
along the slit, i.e., at slit positions of 64.7\arcsec\ (orange) and
60.6\arcsec\ (purple) in Fig.~\ref{fit1} and \ref{fit2} (upper). As
predicted by the explosive evaporation, the coronal line of \fexxi
exhibits the blue shifts, while the chromospheric line of \ci
displays the red shifts at the same interval. \fexxi increases its
blue shifts rapidly to the maximum, then gradually and monotonically
decreases to zero, continuously turns towards the red shifts. There
are two possible explanations for these red shifts. Firstly, they
maybe due to the hot material falling to the chromosphere along the
flare loop after cooling. Secondly, they could be the signatures of
loop contracting as seen in many imaging observations of flare
arcades
\citep[e.g.,][]{Wang92,Ambastha93,Sui03,Li05,Li06,Ji06,Ji07,Zhou09,Liu13,Ning13,Yan13,Zhou13,Kushwaha15,Wang15}.
Similar to \fexxi, \ci increases its red shifts firstly to the
maximum and then decreases to a stable state of about 24 km s
$^{-1}$. There are two different physical scenarios to explain \ci
red shifts. The explosive peak of \ci red shifts could be due to
chromospheric condensation corresponding the evaporation detected as
\fexxi blue shifts at the same intervals, while the red shifts of
\ci on the decay phase could be due to the material falling back to
chromosphere or the loop contracting, which results into the red
shifts, not explosive but stable. \fexxi and \ci show explosive peak
at the Doppler velocities, and the peak values can reach about -200
km s$^{-1}$ and 27 km s$^{-1}$, respectively. Although the peak time
is different from the positions on the slit, the Doppler velocities
exhibit the similar explosive peak. This is because the flare ribbon
expands with time. The dashed lines in Fig.~\ref{light} (c) give the
three times of the standard deviation (3$\sigma$) of the Doppler
velocities from the quiet intervals (black profiles). The pluses
(`+') mark the points where the speed values above 3$\sigma$ and
corresponding to the HXR peaks. Here, the evaporation time scale can
be estimated from \fexxi blue shifts. It is about 10 minutes for the
2014 September 10 flare, which is consistent with recent findings
for the same flare \citep{Graham15,Tian15}. The pluses in
Fig.~\ref{light} (e) mark the same points as that in
Fig.~\ref{light} (c). After the blue shifts, \fexxi becomes red
shifts with a value of $\sim$24 km s$^{-1}$. Meanwhile, \ci red
shift velocity has the similar value ($\sim$24 km s$^{-1}$),
indicating the material falling downward with a similar speed from
the corona. The error bars of the multi-Gaussian fitting are
displayed every 20-point with 2-$\delta$ uncertainties in
Fig.~\ref{light} (c, e). The orange and purple colors are for two
different positions on the slit, respectively. On the flare ribbon,
the enhancement emissions result into the fitting speed errors
decreasing to about 2 km s$^{-1}$ (i.e., $\delta$ = $\sim$2 km
s$^{-1}$). Same as Fig.~\ref{light} (a, c, and e), Fig.~\ref{light}
(b, d, and f) shows the light curves of X-ray, \fexxi, and \ci
Doppler velocities for 2014 October 22 flare, which is well detected
by {\it RHESSI}. As mentioned before, this event is detected by {\it
IRIS} in `raster' mode, and Fig.~\ref{light} (d and f) shows the
Doppler velocities at the position of the second step in each
raster. Same as Fig.~\ref{vel2}, the time resolution is as low as
131 s. And the evaporation time scale of $\sim$11 minutes is
estimated from \fexxi blue shifts. The error bars of the
multi-Gaussian fitting are shown by every 2-point for each Doppler
velocity.

Fig.~\ref{light} (a) shows that there are three impulsive HXR peaks
($\geq$ 27.3 keV) as marked by `1', `2', and `3'. It is clear that
the last two HXR peaks (`2' and `3') are well correlated with the
Doppler velocity peaks at two distinct positions, as shown in
Fig.~\ref{light} (c, e), whatever \fexxi or \ci. In other words,
both \fexxi and \ci exhibit a `increase-peak-decrease' pattern of
their Doppler velocities, well correlated with the
`rising-maximum-decay' phase at HXR emission. Considering the
velocity direction, HXR light curves are anti-correlated with \fexxi
Doppler velocities, while correlated with \ci Doppler velocities.
This situation is well seen in the 2014 October 22 flare as well.
There are two HXR peaks at the channel of 25$-$50 keV, as marked by
`1' and `2' in Fig.~\ref{light} (b). They are well correlated with
the \fexxi and \ci Doppler velocity peaks at two distinct positions,
as shown in Fig.~\ref{light} (d, f).

Fig.~\ref{scatter1} plots HXR peaks at 27.3$-$50.9 (or 25$-$50) keV
dependence on the \fexxi and \ci Doppler velocities for the 2014
September 10 and 2014 October 22 flares, respectively. As expected
from the electron-driven evaporation model, we find an
anti-correlation between HXR emissions and coronal line (\fexxi)
evaporation velocities, while correlation between HXR emissions and
chromospheric line (\ci) condensation velocities. The correlation
coefficient above 0.7 indicates that the non-thermal electrons cause
the HXR emissions and drive the explosive evaporation simultaneously
after precipitating in the chromosphere. Fig.~\ref{scatter1} (c, d)
shows only 4 points used for the correlation of each HXR peak due to
the low time resolution of {\it IRIS} raster for the 2014 October 22
flare. For the peak `1', the HXR emission at 25$-$50 keV starts an
intensity as small as about 20 counts s$^{-1}$, then becomes more
than one order bigger after the maximum. That results into a single
point at about 20 counts s$^{-1}$ of HXR emission in
Fig.~\ref{scatter1} (c, d). The correlation coefficient will become
larger if this point is omitted.

Fig.~\ref{scatter2} plots the two HXR peaks at higher energy
channels (such as 50.9$-$102.3 keV, and 102.3$-$296.4 keV)
dependence on \fexxi and \ci Doppler velocities for the 2014
September 10 flare. The similar anti-correlation between HXR
emissions and coronal line (\fexxi) Doppler shifts, while
correlation between HXR emissions and chromospheric line (\ci)
Doppler shifts are found during the same interval. The higher
correlation coefficients ($>$ 0.85) further confirm our results.

\section{Discussions}
Using {\it IRIS} spectral observations on the flare ribbon and HXR
observations from {\it Fermi} or {\it RHESSI}, we investigate the
relationship between HXR emissions and Doppler velocities during the
explosive evaporations in two X-class solar flares on 2014 September
10 and October 22. Using the multi-Gaussian fitting, Doppler
velocities of \fexxi and \ci are detected from {\it IRIS} spectral
observations. At certain position on the slit, \fexxi and \ci
display their Doppler velocities with a `increase-peak-decrease'
pattern, which is well related to the `rising-maximum-decay' phase
of HXR emissions. Consistent with previous findings
\citep{Fisher85a,Fisher85b,Milligan09,Brosius13,Tian14}, we find a
high anti-correlation between HXR emissions and coronal line Doppler
shifts of \fexxi, and a high correlation between HXR emissions and
chromospheric line Doppler shifts of \ci, indicating the electron
beam-driven explosive evaporation in these two solar flares. The
similar results are also found in the recent paper by
\citet{Tian15}, who get a high correlation between \fexxi blue
shifts and derivative of {\it GOES} SXR for the 2014 September 6 and
10 flares.

\begin{figure}
\epsscale{1.0} \plotone{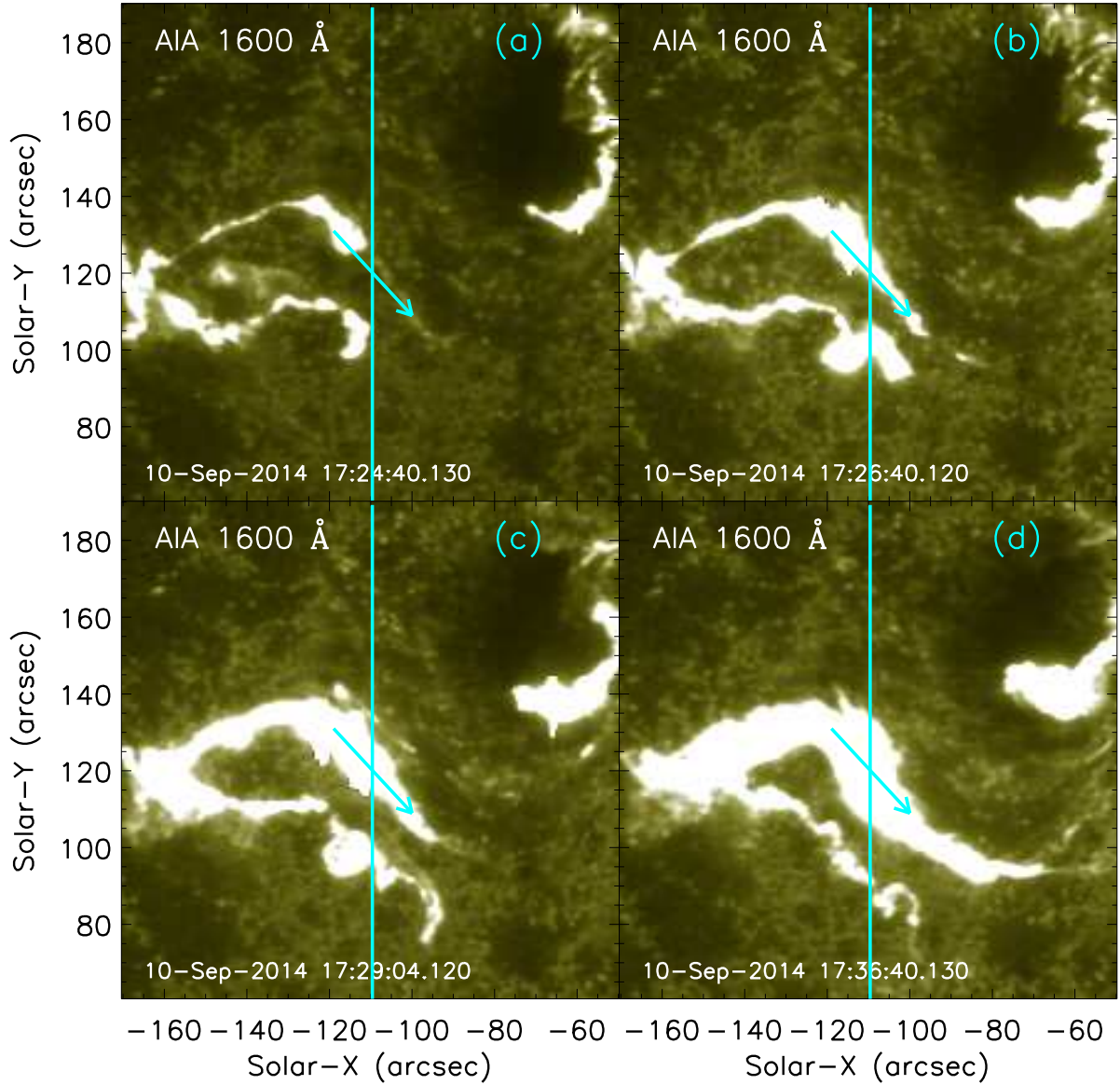} \caption{Time sequences of AIA 1600
{\AA} images for the 2014 September 10 flare. The vertical line
represents the {\it IRIS} slit positions, and the arrows mark the
ribbon propagation. \label{ribbon}}
\end{figure}

\begin{figure}
\epsscale{1.0} \plotone{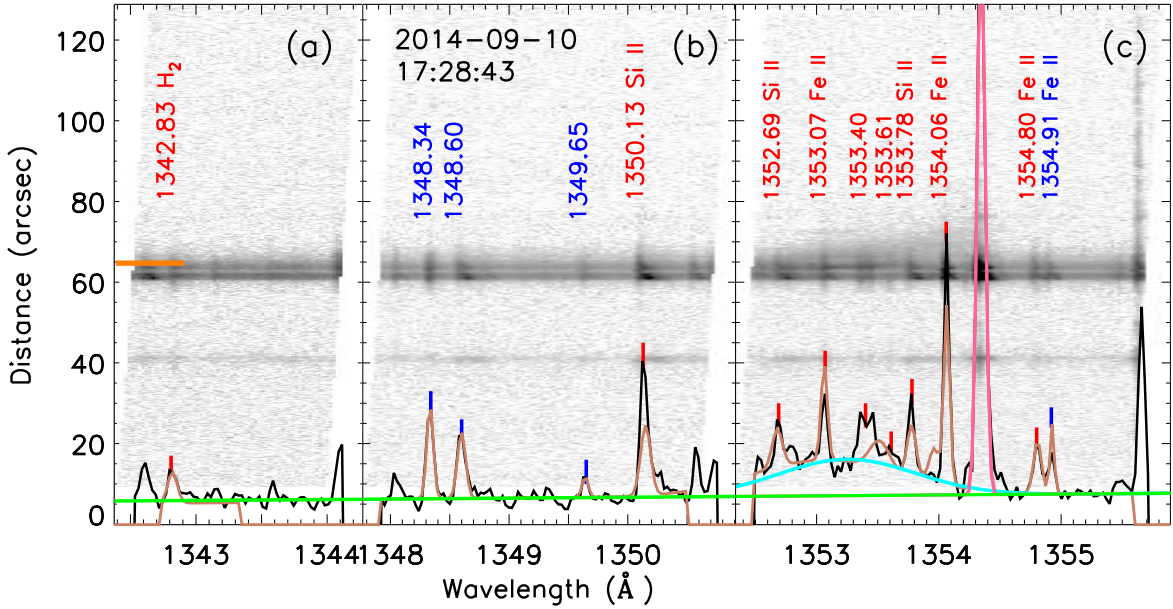} \caption{Three {\it IRIS} spectra
windows ((a) for `1343', (b) for `\fexii', and (c) for `\oi') at
17:28:43 UT for 2014 September 10 flare. The black profiles are
detected at $\sim$64.7\arcsec\ along the slit positions (marked by
the short horizontal line). The brown profiles represent the
multi-Gaussian fitting. The turquoise profile is \fexxi, the magenta
profile is \ci, and the green is the background. The other 13
emission lines used in this paper are labeled by vertical ticks.
\label{fit1}}
\end{figure}

\begin{figure}
\epsscale{1.0} \plotone{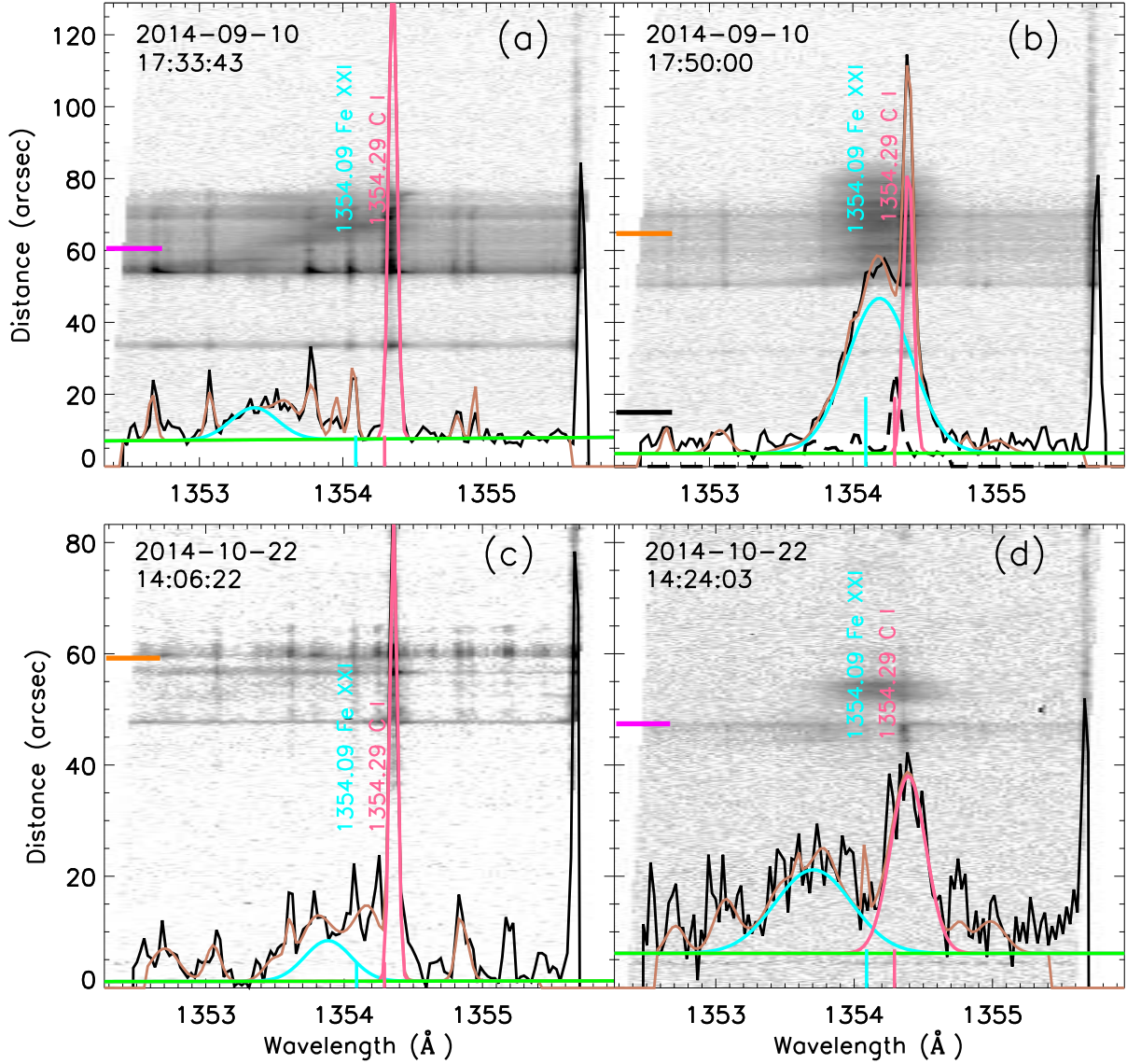} \caption{Similar as Fig.~\ref{fit1}
(c), {\it IRIS} flare spectra at `\oi' window at four time on 2014
September 10 and October 22, respectively. The black profiles are
detected at various positions on the slit marked by the short
horizontal lines. The brown profile represents the spectral fitting.
The turquoise profile is \fexxi, the magenta profile is \ci, and the
green is the background. The short vertical lines mark the rest
wavelengths of \fexxi (turquoise) and \ci (magenta), respectively.
The dashed black profile is the spectra from the non-flaring region.
\label{fit2}}
\end{figure}

\begin{figure}
\epsscale{1.0} \plotone{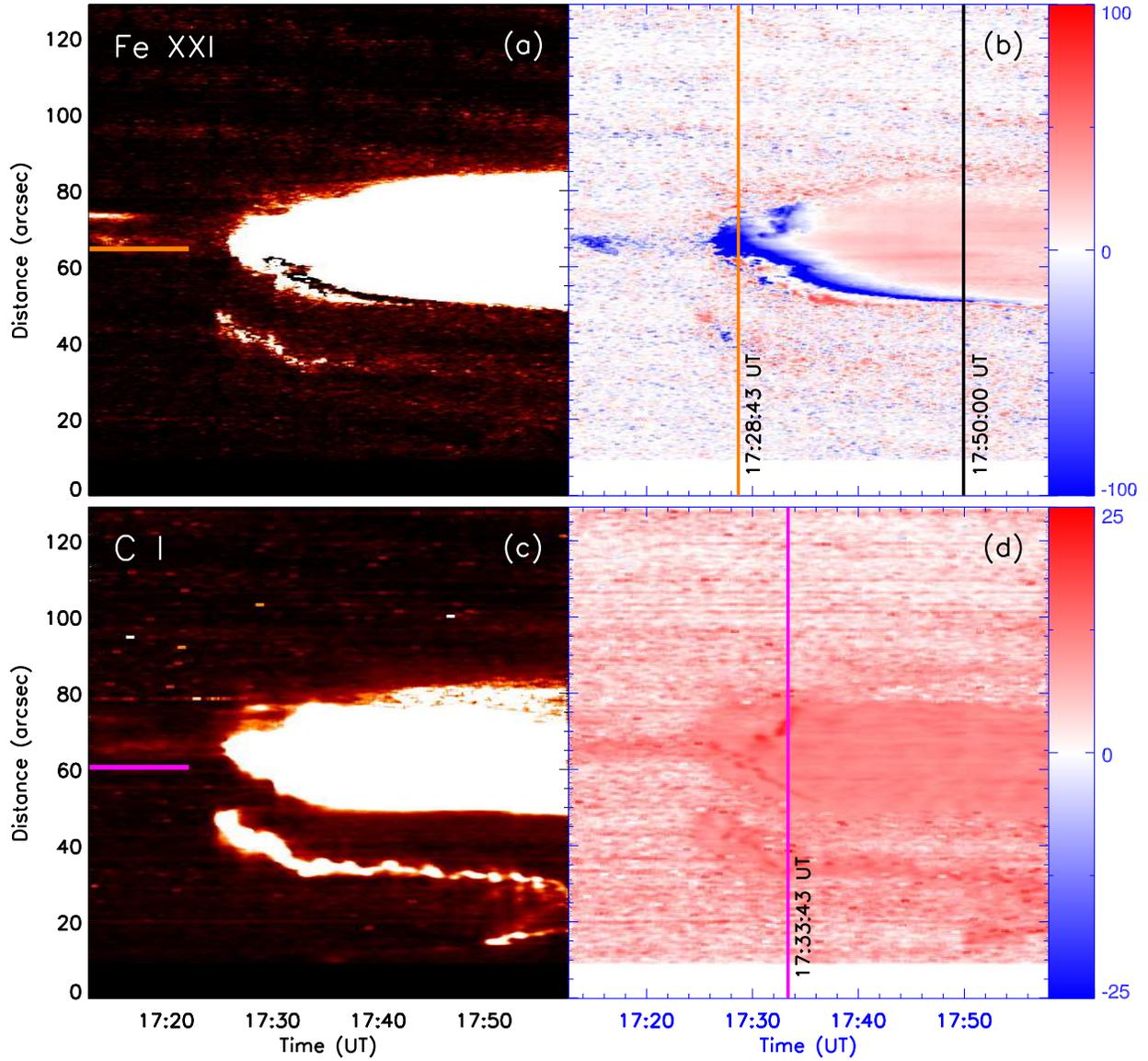} \caption{The space-time diagrams of
intensities and Doppler velocities of \fexxi and \ci for the 2014
September 10 flare. Y-axis is the distance along the whole {\it
IRIS} slit. The spectral profiles at two positions on the slit
marked by short horizontal lines and three times marked by vertical
lines are shown in Figs.~\ref{fit1} and \ref{fit2} (a, b)
\label{vel1}}
\end{figure}

As noted earlier, Fig.~\ref{light} plots the \fexxi and \ci Doppler
velocities at two distinct positions respectively, which results
into a good correlation with HXR peaks simultaneously. The peak time
of Doppler velocities would be changed for various positions on the
slit. In other words, Doppler velocities on the other positions are
not well correlated with the HXR peaks at the same intervals.
However, the time profiles of Doppler velocities at any positions
exhibit the similar shape as shown in Fig.~\ref{light}. We can still
obtain the high correlations between Doppler velocities and HXR
peaks after shifting an interval of Doppler velocity peaks. These
intervals are different for the various positions along the slit,
and it is an open question that a temporal correlation does not
exist at these positions along the slit. Fig.~\ref{light} just gives
the Doppler velocities at two special positions on the slit. They
don't need to shift with time to do the correlation in
Figs.~\ref{scatter1} and~\ref{scatter2}. For example, the first peak
of the purple curve in Fig.~\ref{light} (e) is at 17:29 UT, and it
could be correspond well with the peak `2' in {\it Fermi} data if it
is shifted an interval of about 1 minutes. While the second peak at
17:33 UT are corresponding well with the HXR peak `3' without
shifting. \citet{Tian15} reported a time delay ($\sim$0.5$-$2.0
minutes) for the larger correlation between the \fexxi blue shifts
and SXR derivative on 2014 September 10 flare. This delay should be
caused by the spectral profiles at different positions along the
{\it IRIS} slit. When the flare ribbon propagates toward the
South-West and expands with time, the peak of \fexxi blue shifts
changes with the slit positions, as well as with the time (seen in
Figs.~\ref{ribbon} and \ref{vel1}). In this paper, we fit all the
spectra along the {\it IRIS} slit at all time to obtain the
space-time images of Doppler velocities, and get the time evolution
of Doppler velocities at any positions on the slit. On the other
hand, both \fexxi and \ci exhibit red shifts on the flare decay
phase, especially for the 2014 September 10 flare. As the cross
(`$\times$') marked in Fig~\ref{light} (c, e), \fexxi has a velocity
of $\sim$24 km s$^{-1}$, while \ci has a small peak. The spectral
profile at this time is given in Fig.~\ref{fit2} (b), which shows
strong emissions at \fexxi and \ci. At this position, \fexxi and \ci
have line centers of about 1354.19 and 1354.40 {\AA}, line widths of
524.33 m{\AA} and 76.96 m{\AA}. They show red shifts from their rest
wavelengths. These red shifts of \fexxi and \ci could be caused by
the material falling down or the loop contracting at the decay phase
of the flares.

\begin{figure}
\epsscale{1.0} \plotone{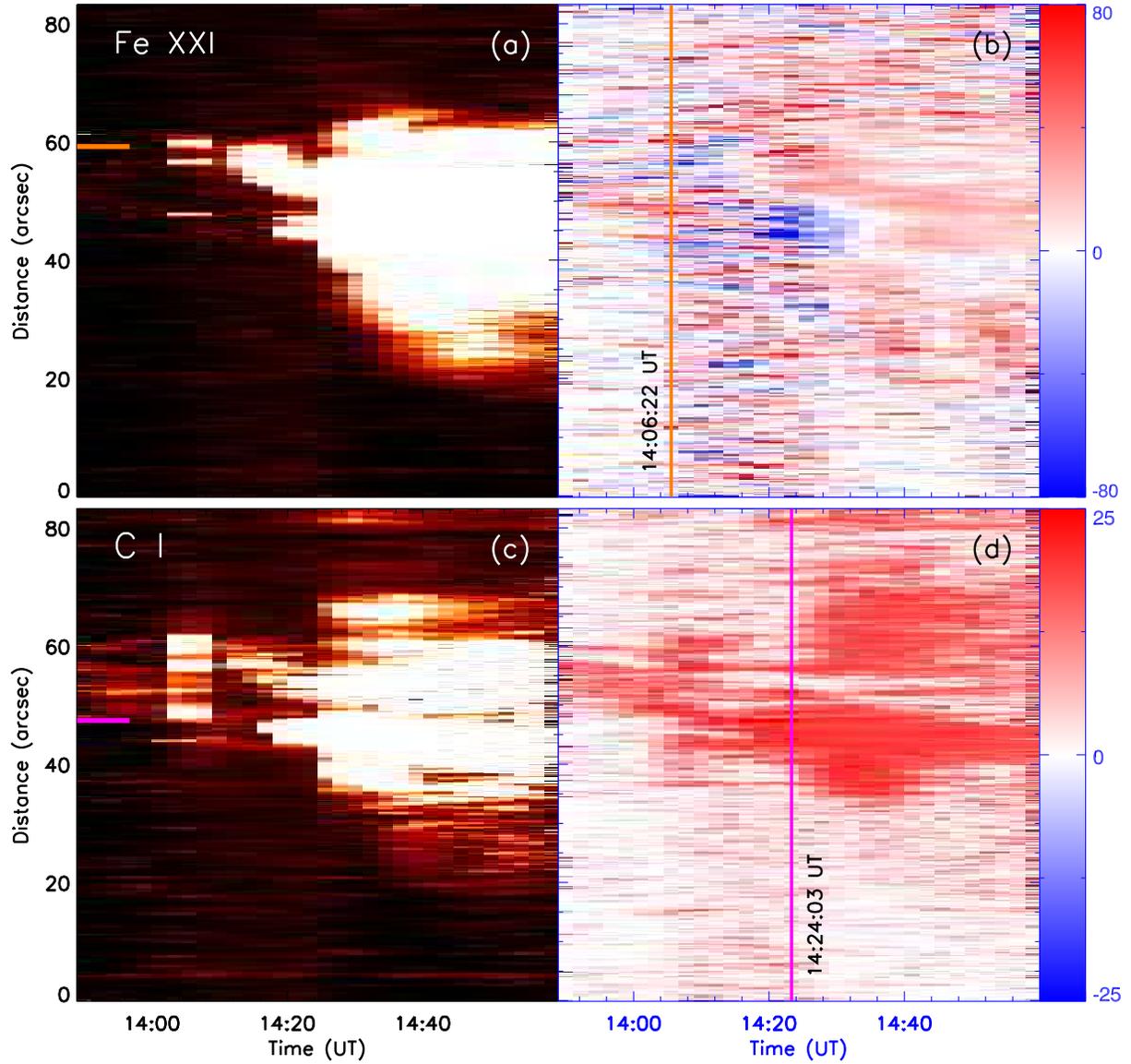} \caption{Same as Fig.~\ref{vel1},
but for the 2014 October 22 flare. Y-axis gives the distance between
two short blue lines along the {\it IRIS} slit in Fig.~\ref{image}
(c, d). The spectral profiles at two positions on the slit marked by
short horizontal lines and two times marked by vertical lines are
given in Fig.~\ref{fit2} (c, d). \label{vel2}}
\end{figure}

\begin{figure}
\epsscale{1.0} \plotone{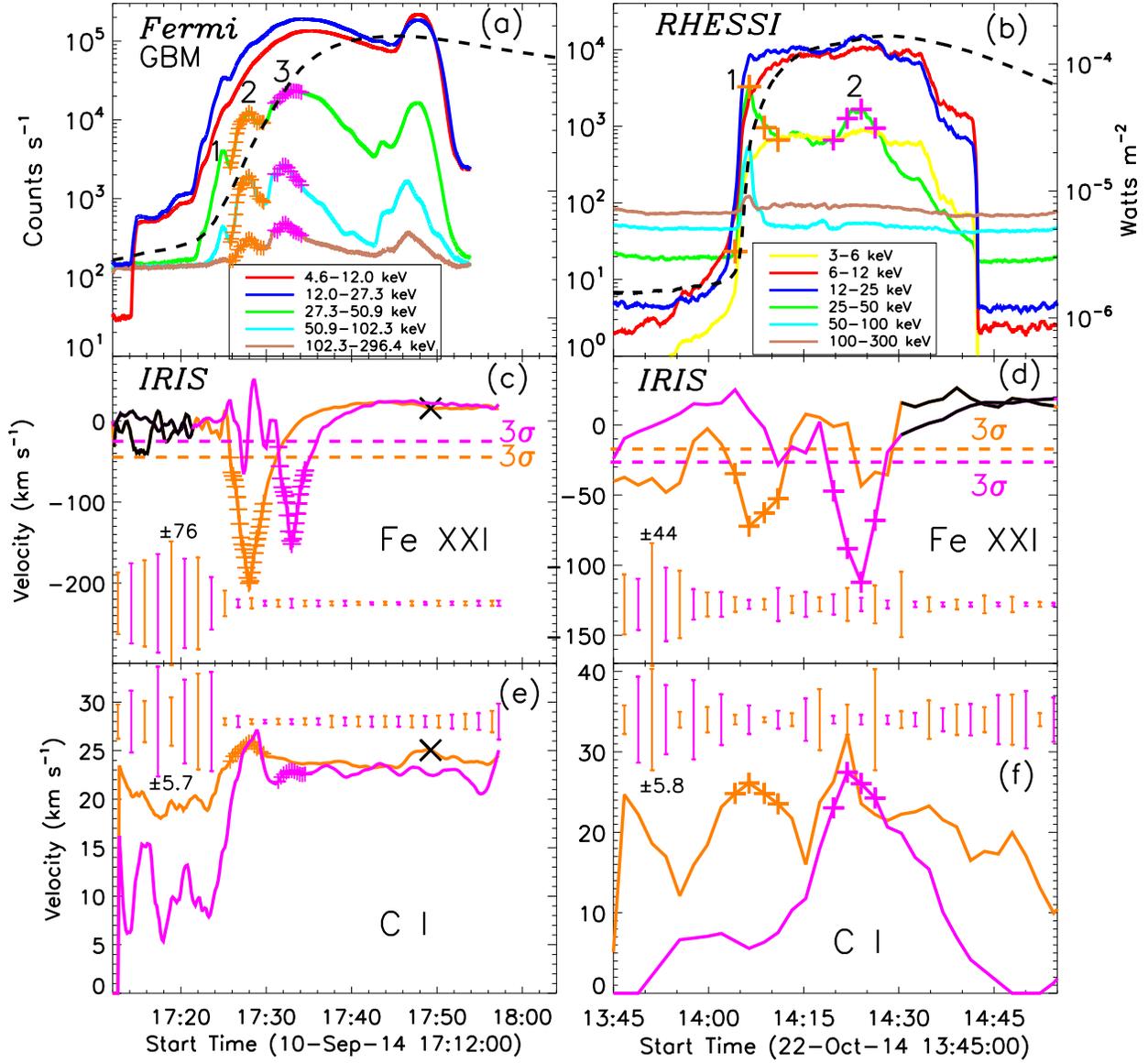} \caption{Top: the light curves from
{\it Fermi} for the 2014 September 10 flare (a) and from {\it
RHESSI} for the 2014 October 22 flare (b), and from {\it GOES}
(dashed black lines) for both events. Middle and Bottom: the
detected Doppler velocities of \fexxi and \ci at two distinct
positions on the slits (seen text in details). The dashed lines are
the three times of the standard deviation, the pluses (`+') mark the
points with speed values above 3$\sigma$ and corresponding HXR
peaks. The cross (`$\times$') marks the point in the decay phases of
the flare, and its spectral profile is given in Fig.~\ref{fit2} (b).
The error bars represent the 2-$\delta$ uncertainties from the
multi-Gaussian fitting in the middle and bottom panels.
\label{light}}
\end{figure}

\begin{figure}
\epsscale{1.0} \plotone{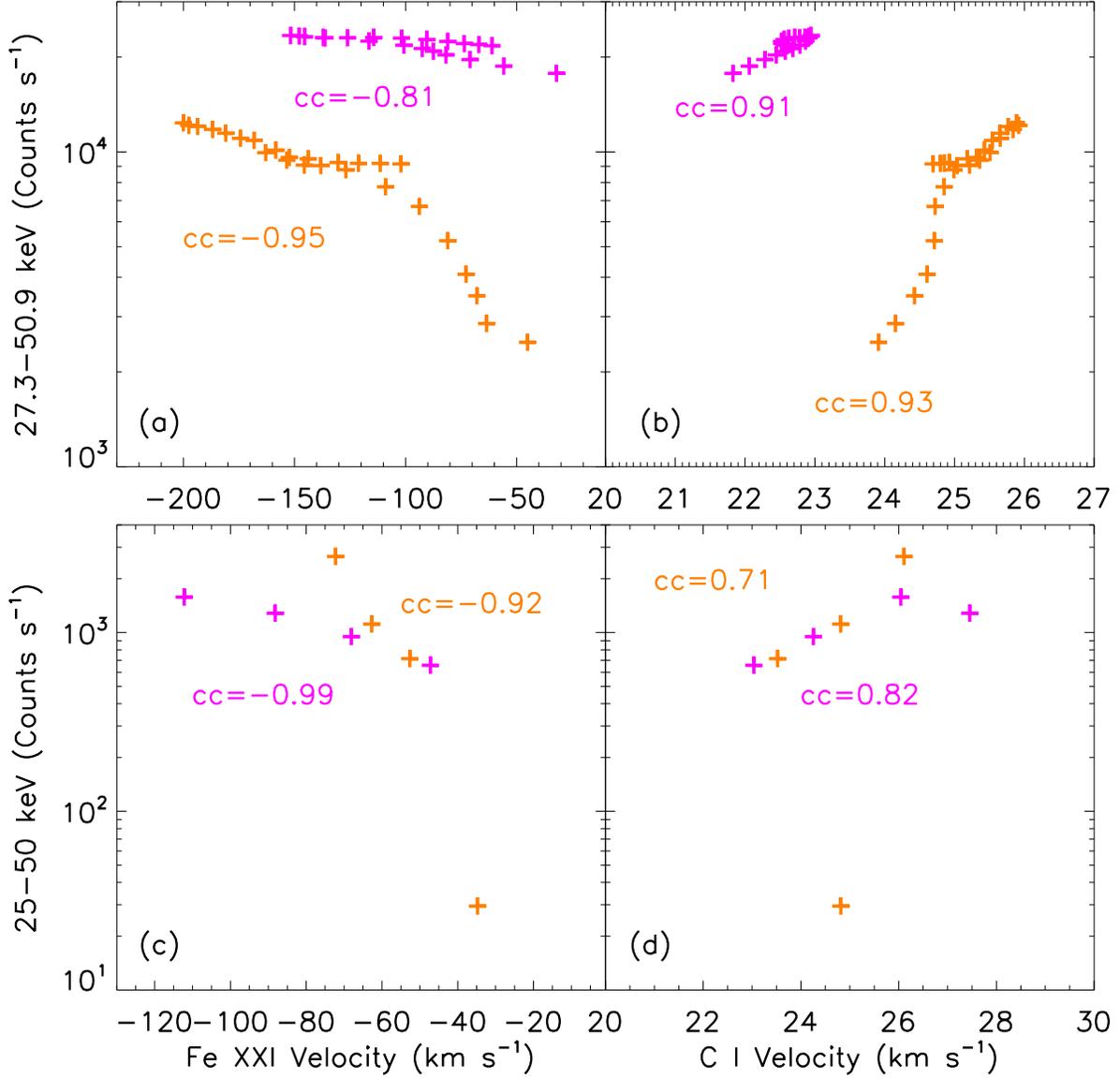} \caption{Scatter plots of two HXR
peaks (orange and purple) dependence on Doppler velocities of \fexxi
and \ci for the 2014 September 10 (a, b) and October 22 flares (c,
d). The correlation coefficients (cc) are given. \label{scatter1}}
\end{figure}

\begin{figure}
\epsscale{1.0} \plotone{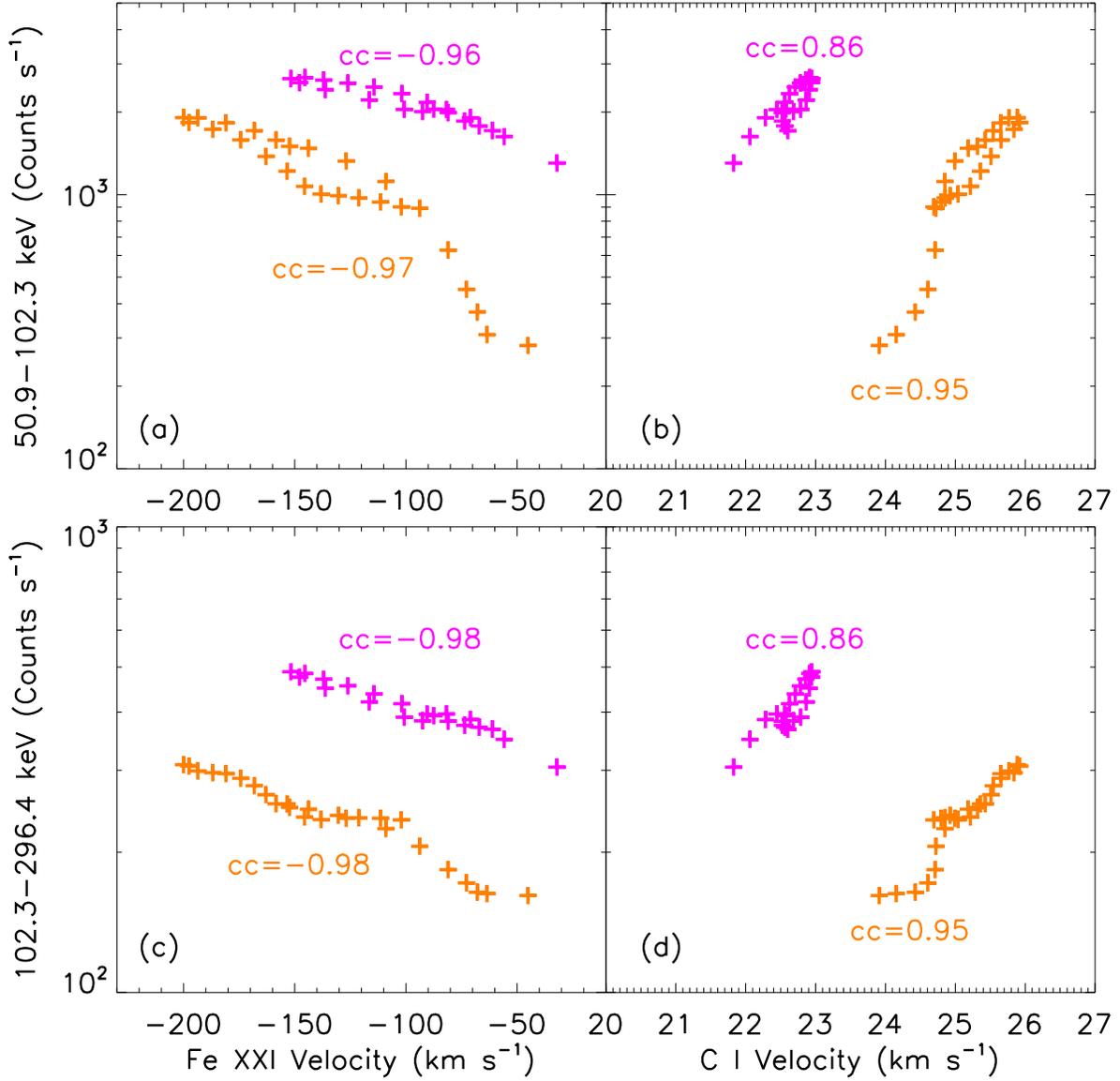} \caption{Same as
Fig.~\ref{scatter1}, but HXR emissions at 50.9$-$102.3 keV and
102.3$-$296.4 keV for 2014 September 10 flare. \label{scatter2}}
\end{figure}

The rest wavelength of chromospheric line \ci is set as 1354.29
{\AA}, which is detected from the quiet Sun (seen in Fig.~\ref{fit2}
(b)). This is consistent with recent studies from {\it IRIS}
observations \citep{Polito15,Sadykov15}. However, as noted earlier,
\fexxi is a hot coronal line, which is absent on the quiet Sun.
Therefore, we can not determine the rest wavelength of \fexxi from
non-flaring spectrum. There are different line centers of \fexxi in
the literatures. The first value of 1354.1 {\AA} for \fexxi has been
identified from the spectra of solar flares by \citet{Doschek75}.
Then the rest wavelength of \fexxi is though between 1354.06 {\AA}
and 1354.12 {\AA}
\citep[e.g.,][]{Cheng79,Mason86,Feldman00,Innes03a,Innes03b,Wang03}.
In this paper, the rest wavelength of \fexxi is set to 1354.09
{\AA}, and this value is similar as that in recent studies
\citep{Tian14,Tian15,Young15,Graham15,Polito15,Sadykov15} about
\fexxi from {\it IRIS} spectral data. Considering the broad range of
\fexxi, the rest wavelength has an uncertainty of about $\pm$0.03
{\AA}, corresponding the Doppler velocity of about $\pm$6.6 km
s$^{-1}$. The rest wavelength of \fexxi is probably taken from the
decay phase of the flare due to it's stable emission, as the example
of the 2014 September 10 flare. However, the wavelength value is
$\sim$1354.19 {\AA} (i.e., the cross (`$\times$') in
Fig.~\ref{light} (c, d)) at the decay phase of this flare, which is
much bigger than the rest wavelength in previous studies. It is a
fact that the coronal line of \fexxi is absent in the non-flare
regions where the temperature is not hot enough. Therefore, the
intensities and Doppler velocities of \fexxi in Figs.~\ref{vel1} (a,
b) and \ref{vel2} (a, b) outside the flare regions are invalid, they
are fitting noises from the observation data. Same as the
chromospheric line of \ci, its Doppler velocities (Figs.~\ref{vel1}
(d) and \ref{vel2} (d)) outside the flare regions should also be
fitting noises from the observation data.

The multi-Gaussian fitting is used to determine the Doppler
velocities of \fexxi and \ci in this paper. There are three sources
of uncertainties to their Doppler shifts. Firstly, as shown in
Fig.~\ref{light}, the fitting errors are very large in the non-flare
regions and become much less in the flare regions, no matter \fexxi
or \ci. This error is mathematic from the fitting method. Secondly,
the red wing enhancement or redshifted component of the cool lines
(i.e., \feii, \siii, and \ci) could affect the identification of
\fexxi \citep{Young15,Tian15}. There is no way to exactly rule out
these blending lines from \fexxi at present time. Using tied
intensities, line centers and widths could be a better approach as
normally lines from the same ion or similar lines should have
similar behaviors. Therefore, we constrained and tied these lines
with the emission lines in other windows (seen in Table~\ref{table})
to eliminate the influence of these cool lines. Thirdly, the
asymmetries of \ci line should also affect the derived red shift of
\ci. Because the Gaussian fitting of such asymmetrical lines tends
to underestimate the velocity of chromospheric condensation,
especially for the optically thick lines, i.e., \mgii line
\citep{Graham15}, and H$\alpha$ line \citep{Ding95}. A good way to
determine the Doppler shifts from the asymmetric line is the
bisector method \citep[i.e.,][]{Ding95,Graham15}, which could
accurately determine the Doppler shift, i.e., especially the red
shift for the asymmetrical \ci in our case. However, not the
bisector, but the Gaussian fitting is used in this paper. Because
\ci is one of the multi-Gaussian fitting in our code, while the
bisector method is a better way for the isolated line.

In this paper, we find the blue shifts of \fexxi quickly increases
from zero to the maximum (more than 200 km s$^{-1}$) before
decreasing, which is different from \citet{Polito15} finding that
the evaporation speed shows a monotonic decrease once appearing at
the flare ribbon. This would be because we take the whole history of
the \fexxi Doppler velocities during the impulsive phase of the
flare, while \citet{Polito15} only show the Doppler velocities after
its maximum. On the other hand, their data is in `raster' mode,
rather than `sit and stare' mode. The time resolution is not high.
The 2014 October 22 flare in this paper is in `raster' mode, and has
a lower time resolution than the 2014 September 10 flare. But we can
also detect the Doppler velocity increase before its maximum. The
flare on 2014 September 10 is also studied by
\citet{Graham15,Tian15}. Both of them found very clearly monotonic
decrease of the \fexxi blue shifts, but no increase before the peak,
as shown in Fig.~13 in \citet{Tian15}. They also found the
evaporation within about 9 minutes, from 17:32 UT to 17:41 UT, which
is similar to our finding of $\sim$10 minutes evaporation. From the
orange curves in Fig.~\ref{light} (c), the start time of evaporation
is about 17:26 UT (velocity above 3$\sigma$), and the end time is
about 17:36 UT (velocity equally to zero), while the peak time is
about 17:28 UT. \citet {Tian15} showed the monotonic decrease of
\fexxi blue shifts for the 2014 September 10 flare after the maximum
at 17:32 UT. Their velocity curve is at position around 117.8\arcsec
on the slit. However, we give all the histories of the flare at the
position of $\sim$64.7\arcsec along the slit in Fig.~\ref{light}
(c), i.e., from the flare onset at 17:21 UT. On the other hand,
\citet{Tian15} found the same evaporation pattern of
`increase-peak-decrease' for 2014 September 6 flare when their
fitting results cover the whole histories of the flare. In fact,
previous studies had been reported the similar evolution of Doppler
velocities from hot coronal lines (i.e., \fexii, \fexix, \fexxi, and
so on) in the solar flares, such as the evolution of
`increase-peak-decrease' pattern
\citep[e.g.,][]{Wang03,Kamio05,Brosius09,Raftery09,Li14,Tian15}.
Fig.~\ref{light} (e) also shows that the chromospheric \ci line
seems to display several small peaks after its maximum. They could
be related to the HXR emissions at the decay phase or the materials
with various speeds falling back to chromosphere. Meanwhile, they
probably are the spectral signatures of the various loop
contracting.

\acknowledgments We would like to thank two anonymous referees for
their valuable comments to improve the manuscript. The data used in
this paper are from {\it IRIS}, {\it Fermi}, {\it RHESSI}, {\it
GOES}, and {\it SDO}. {\it IRIS} is a NASA small explorer mission
developed and operated by LMSAL with mission operations executed at
NASA Ames Research center and major contributions to downlink
communications funded by the Norwegian Space Center (NSC, Norway)
through an ESA PRODEX contract. This study is supported by NSFC
under grants 11203083, 11333009, 11303101, 11473071, 11573072, 973
program (2011CB811400, 2014CB744200) and Laboratory NO.
2010DP173032. We are grateful to Dr. D.~E. Innes for helping to
develop the spectral fitting code of {\it IRIS} data. The authors
would thank to Professors H.~M. Wang and H.~S. Ji for improving the
paper.

\begin{table} \caption{The parameters of 15 emission lines at three {\it IRIS}
spectral windows.} \centering
\begin{tabular}{c c c c c}
\hline\hline
{\it IRIS} window   & Ion     & Wavelength  ({\AA})  &  Width (m{\AA}) &  Intensity tied to  \\
\hline
                  &\siii    & 1352.69$\pm$0.102   &  $\leq$ 260   &   \siii 1350.13     \\
                  &\feii    & 1353.07$\pm$0.051   &  $\leq$ 88    &   \feii 1354.80     \\
                  &Unknown  & 1353.40$\pm$0.061   &  $\leq$ 102   &   H$_2$ 1342.83     \\
                  &Unknown  & 1353.61$\pm$0.061   &  $\leq$ 102   &   H$_2$ 1342.83     \\
`\oi'             &\siii    & 1353.78$\pm$0.102   &  $\leq$ 260   &   \siii 1350.13     \\
                  &\feii    & 1354.06$\pm$0.051   &  $\leq$ 88    &   \feii 1354.80     \\
                  &\fexxi   & 1354.09$\pm$1.28    &  $\geq$ 230   &                     \\
                  &\ci      & 1354.29$\pm$0.26    &  $\leq$ 130   &                     \\
                  &\feii    &  1354.80$\pm$0.051   &  $\leq$ 88   &                     \\
                  &\feii    &  1354.91$\pm$0.061   &  $\leq$ 102  &                     \\
\hline
                  &\siii    &  1350.13$\pm$0.102   &  $\leq$ 260  &                     \\
                  &Unknown  &  1348.34$\pm$0.067   &  $\leq$ 102  &                     \\
`\fexii'          &Unknown  &  1348.60$\pm$0.067   &  $\leq$ 102  &                     \\
                  &Unknown  &  1349.65$\pm$0.051   &  $\leq$ 77   &                     \\
\hline
`1343'            &H$_2$    &  1342.83$\pm$0.061   &  $\leq$ 102  &                     \\

\hline\hline
\end{tabular}
\label{table}
\end{table}

\begin{thebibliography}{}
\bibitem[Acton et al.(1982)]{Acton82} Acton, L.~W., \& Leibacher, J.~W., \& Canfield, R.~C., et al. \ 1982, \apj, 263, 409
\bibitem[Ambastha et al.(1993)]{Ambastha93} Ambastha, A., Hagyard, M.~J., \& West, E.~A.\ 1993, \solphys, 148, 277
\bibitem[Antonucci et al.(1982)]{Antonucci82} Antonucci, E, \& Gabriel, A.~H., \& Acton, L.~W., et al. \ 1982, \solphys, 78, 107
\bibitem[Aschwanden \& Benz(1995)]{Aschwanden95} Aschwanden, M.~J., \& Benz, A.~O. \ 1995, \apj, 438, 997
\bibitem[Berlicki et al.(2005)]{Berlicki05} Berlicki, A., \& Heinzel, P., \& Schmieder, B., et al. \ 2005, \aap, 430, 679
\bibitem[Brosius \& Holman (2007)]{Brosius07} Brosius, J.~W., \& Holman, G.~D. \ 2007, \apjl, 659, L73
\bibitem[Brosius(2009)]{Brosius09} Brosius, J.~W. \ 2009, \apj, 701, 1209
\bibitem[Brosius(2013)]{Brosius13} Brosius, J.~W. \ 2013, \apj, 762, 133
\bibitem[Brown(1971)]{Brown71} Brown, J.~C. \ 1971, \solphys, 18, 489
\bibitem[Brown(1973)]{Brown73} Brown, J.~C. \ 1973, \solphys, 31, 143
\bibitem[Carmichael(1964)]{Carmichael64} Carmichael, H.\ 1964, NASA Special Publication, 50, 451
\bibitem[Chen \& Ding(2010)]{Chen10} Chen, F., \& Ding, M.~D. \ 2010, \apj, 724, 640
\bibitem[Cheng et al.(1979)]{Cheng79} Cheng, C.~C., \& Feldman, U., \& Doschek, G.~A.\ 1979, \apj, 233, 736
\bibitem[Cheng et al.(2015)]{Cheng15} Cheng, X.,\& Ding, M.~D., \& Fang, C.\ 2015, \apj, 804, 82
\bibitem[Czaykowska(1999)]{Czaykowska99} Czaykowska, A., \& De Pontieu, B., \& Alexander, D., \& Rank, G. \ 1999, \apjl, 521, L75
\bibitem[Curdt et al.(2001)]{Curdt01} Curdt, W., \& Brekke, P., \& Feldman, U., et al.\ 2001, \aap, 375, 591
\bibitem[Curdt et al.(2004)]{Curdt04} Curdt, W., \& Landi, E., \& Feldman, U.\ 2004, \aap, 427, 1045
\bibitem[De Pontieu et al.(2014)]{Dep14} De Pontieu, B., \& Title, A.~M., \& Lemen, J.~R., et al. \ 2014, \solphys, 289, 2733
\bibitem[Del Zanna et al.(2006)]{Del06} Del Zanna, G., \& Schmieder, B., \& Mason, H., \& Berlicki, A., \& Bradshaw, S. \ 2006, \solphys, 234, 95
\bibitem[Ding et al.(1995)]{Ding95} Ding, M.~D., \& Fang, C., \& Huang, Y.~R.\ 1995, \solphys, 158, 81
\bibitem[Ding et al.(1996)]{Ding96} Ding, M.~D., \& Watanabe, T., \& Shibata, K., et al. \ 1996, \apj, 458, 391
\bibitem[Doschek et al.(1975)]{Doschek75} Doschek, G.~A., \& Dere, K.~P., \& Sandlin, G.~D., et al.\ 1975, \apjl, 196, L83
\bibitem[Doschek et al.(1980)]{Doschek80} Doschek, G.~A., \& Feldman, U., \& Kreplin, R.~W., \& Cohen, L. \ 1980, \apj, 239, 725
\bibitem[Doschek et al.(2013)]{Doschek13} Doschek, G.~A., \& Warren, H.~P., \& Young, P.~R. \ 2013, \apj, 767, 55
\bibitem[Emslie et al.(1978)]{Emslie78} Emslie, A.~G., \& Brown, J.~C., \& Donnelly, R.~F.\ 1978, \solphys, 57, 175
\bibitem[Falewicz et al.(2009)]{Falewicz09} Falewicz, R., \& Rudawy, P., \& Siarkowski, M. \ 2009, \aap, 508, 971
\bibitem[Feldman et al.(1980)]{Feldman80} Feldman, U., \& Doschek, G.~A., \& Kreplin, R.~W., \& Mariska, J.~T. \ 1980, \apj, 241, 1175
\bibitem[Feldman et al.(2000)]{Feldman00} Feldman, U., \& Curdt, W., \& Landi, E., \& Wilhelm, K.\ 2000, \apj, 544, 508
\bibitem[Fisher et al.(1985a)]{Fisher85a} Fisher, G.~H, \& Canfield, R.~C., \& McClymont, A.~N. \ 1985a, \apj, 289, 414
\bibitem[Fisher et al.(1985b)]{Fisher85b} Fisher, G.~H, \& Canfield, R.~C., \& McClymont, A.~N. \ 1985b, \apj, 289, 425
\bibitem[Graham \& Cauzzi(2015)]{Graham15} Graham, D.~R., \& Cauzzi, G.\ 2015, \apjl, 807, L22
\bibitem[Hirayama(1974)]{Hirayama74} Hirayama, T. \ 1974, \solphys, 34, 323
\bibitem[Huang et al.(2014)]{Huang14} Huang, Z., \& Madjarska,M.~S., \& Xia, L., et al.\ 2014, \apj, 797, 88
\bibitem[Innes et al.(2003a)]{Innes03a} Innes, D.~E., \& McKenzie, D.~E., \& Wang, T.~J. \ 2003a, \solphys, 217, 267
\bibitem[Innes et al.(2003b)]{Innes03b} Innes, D.~E., \& McKenzie, D.~E., \& Wang, T.~J. \ 2003b, \solphys, 217, 247
\bibitem[Ji et al.(2006)]{Ji06} Ji, H.~S., \& Huang, G.~L., \& Wang, H.~M., et al.\ 2006, \apjl, 636, L173
\bibitem[Ji et al.(2007)]{Ji07} Ji, H.~S., \& Huang, G.~L., \& Wang, H.~M. \ 2007, \apj, 660, 893
\bibitem[Ji et al.(2008)]{Ji08} Ji, H.~S., \& Wang, H.~M., \& Liu, C., \& Dennis, B.~R. \ 2008, \apj, 680, 734
\bibitem[Kamio et al.(2005)]{Kamio05} Kamio, S., \& Kurokawa, H., \& Brooks, D.~H., \& Kitai, R., \& UeNo, S. \ 2005, \apj, 625, 1027
\bibitem[Karlicky(1998)]{Karlicky98} Karlicky, M. \ 1998, \aap, 338, 1084
\bibitem[Kushwaha et al.(2015)]{Kushwaha15} Kushwaha, U., \& Joshi, B., \& Veronig, A.~M., \& moon, Y.-J.\ 2015, \apj, 807, 101
\bibitem[Kopp \& Pneuman(1976)]{Kopp76} Kopp, R.~A., \& Pneuman, G.~W. \ 1976, \solphys, 50, 85
\bibitem[Lemen et al.(2012)]{Lemen12} Lemen, J.~R., \& Title, A.~M., \& Akin, D.~J., et al. \ 2012, \solphys, 275, 17
\bibitem[Li \& Ding(2004)]{Li04} Li, J.~P., \& Ding, M.~D. \ 2004, \apj, 606, 583
\bibitem[Li \& Ding(2011)]{Li11} Li, Y., \& Ding, M.~D. \ 2011, \apj, 727, 98
\bibitem[Li et al.(2014)]{Li14} Li, Y., \& Qiu, J., \& Ding, M.~D.\ 2014, \apj, 781, 120
\bibitem[Li \& Zhang(2015)]{Liting15} Li, T., \& Zhang, J.\ 2015, \apjl, 804, L8
\bibitem[Li et al.(2015)]{Li15a} Li, D., \& Ning, Z.~J., \& Zhang, Q.~M.\ 2015, \ 2015, \apj, 807, 72
\bibitem[Li \& Gan(2005)]{Li05} Li, Y.~P., \& Gan, W.~Q.\ 2005, \apjl, 629, L137
\bibitem[Li \& Gan(2006)]{Li06} Li, Y.~P., \& Gan, W.~Q.\ 2006, \apjl, 644, L97
\bibitem[Lin et al.(2002)]{Lin02} Lin, R.~P., \& Dennis, B.~R., \& Hurford, G.~J., et al. \ 2002, \solphys, 210, 3
\bibitem[Liu et al.(2006)]{Liu06} Liu, W., \& Liu, S.~M., \& Jiang, Y.~W., et al. \ 2006, \apj, 649, 1124
\bibitem[Liu et al.(2013)]{Liu13} Liu, W., \& Chen, Q., \& Petrosian, V.\ 2013, \apj, 767, 168
\bibitem[Mason et al.(1986)]{Mason86} Mason, H.~E., \& Shine, R.~A., \& Gurman, J.~B., \& Harrison, R.~A.\ 1986, \apj, 309, 435
\bibitem[Meegan et al.(2009)]{Meegan09} Meegan, C., \& Lichti, G., \& Bhat, P.~N., et al.\ 2009, \apj, 702, 791
\bibitem[Milligan et al.(2006a)]{Milligan06a} Milligan, R.~O., \& Gallagher, P.~T., \& Mathioudakis, M., et al.\ 2006a, \apjl, 638, L117
\bibitem[Milligan et al.(2006b)]{Milligan06b} Milligan, R.~O., \& Gallagher, P.~T., \& Mathioudakis, M., et al.\ 2006b, \apjl, 642, L169
\bibitem[Milligan \& Dennis(2009)]{Milligan09} Milligan, R.~O., \& Dennis, B.~R.\ 2009, \apj, 699, 968
\bibitem[Milligan et al.(2014)]{Milligan14} Milligan, R.~O., \& Kerr, G.~S., \& Dennis, B.~R., et al.\ 2014, \apj, 793, 70
\bibitem[Milligan(2015)]{Milligan15} Milligan, R.~O.\ 2015, arXiv:1501.04829
\bibitem[Ning et al.(2009)]{Ning09} Ning, Z.~J., \& Cao, W.~D., \& Huang, J., et al. \ 2009, \apj, 699, 15
\bibitem[Ning \& Cao (2010)]{Ning10} Ning, Z.~J., \& Cao, W.~D. \ 2010, \apj, 717, 1232
\bibitem[Ning(2011)]{Ning11a} Ning, Z.~J. \ 2011 \solphys, 273, 81
\bibitem[Ning \& Cao(2011)]{Ning11b} Ning, Z.~J., \& Cao, W.~D. \ 2011, \solphys, 269, 283
\bibitem[Ning(2013)]{Ning13} Ning, Z~J.\ 2013, \apss, 346, 307
\bibitem[Nitta et al.(2012)]{Nitta12} Nitta, S., \& Imada, S., \& Yamamoto, T.~T. \ 2012, \solphys, 276, 183
\bibitem[Polito et al.(2015)]{Polito15} Polito, V., \& Reeves, K.~K., \& Del Zanna, G., \& Golub, L., \& Mason, H.~E.\ 2015, \apj, 803, 84
\bibitem[Raftery et al.(2009)]{Raftery09} Raftery, \& C.~L., \& Gallagher, P.~T., \& Milligan, R.~O., \& Klimchuk, J. ~A. \ 2009, \aap, 494, 1127
\bibitem[Sadykov et al.(2015)]{Sadykov15} Sadykov, V.~M.,\& Vargas Dominguez, S., \& Kosovichev, A.~G., et al.\ 2015, \apj, 805, 167
\bibitem[Schou et al.(2012)]{Schou12} Schou, J., \& Scherrer, P.~H., \& Bush, R.~I., et al.\ 2012, \solphys, 275, 229
\bibitem[Sturrock(1966)]{Sturrock66} Sturrock, P.~A. \ 1966, \nat, 211, 695
\bibitem[Sui \& Holman(2003)]{Sui03} Sui, L.~H., \& Holman, G.~D.\ 2003, \apjl, 596, L251
\bibitem[Syrovatskii \& Shmeleva(1972)]{Syrovat72} Syrovatskii, S.~I., \& Shmeleva, O.~P.\ 1972, \sovast, 16, 273
\bibitem[Teriaca et al.(2006)]{Teriaca06} Teriaca, L., \& Falchi, A., \& Falciani, R., \& Cauzzi, G., \& Maltagliati, L.\ 2006, \aap, 455, 1123
\bibitem[Tian et al.(2014a)]{Tian14} Tian, H., \& Li, G., \& Reeves, K.~E., et al. \ 2014a, \apjl, 797, L14
\bibitem[Tian et al.(2014b)]{Tian14b} Tian, H., \& DeLuca, E., \& Reeves, K.~K., et al.\ 2014b, \apj, 786, 137
\bibitem[Tian et al.(2015)]{Tian15} Tian, H., \& Young, P.~R., \& Reeves, K.~K., et al.\ 2015, arXiv:1505.02736
\bibitem[Veronig et al.(2010)]{Veronig10} Veronig, A.~M., \& Ryb\'{a}k, J., \& G\"{o}m\"{o}ry, P., et al. \ 2010, \apj, 719, 655
\bibitem[Wang et al.(2003)]{Wang03} Wang, T.~J., \& Solanki, S.~K., \& Curdt, W., et al.\ 2003, \aap, 406, 1105
\bibitem[Wang(1992)]{Wang92} Wang, H.~M. \ 1992, \solphys, 140, 85
\bibitem[Wang \& Liu(2015)]{Wang15} Wang, H.~M., \& Liu, C.\ 2015, Research in Astronomy and Astrophysics, 15, 145
\bibitem[W\"{u}lser et al.(1994)]{Wulser94} W\"{u}lser, J.~P., \& Canfield, R.~C., \& Acton, L.~W., et al. \ 1994, \apj, 424, 459
\bibitem[Yan et al.(2013)]{Yan13} Yan, X.~L., \& Pan, G.~M., \& Liu, J.~H., et al.\ 2013, \aj, 145, 153
\bibitem[Young et al.(2015)]{Young15} Young, P.~R., \& Tian, H., \& Jaeggli, S. \ 2015, \apj, 799, 218
\bibitem[Zhang \& Ji(2013)]{Zhang13} Zhang, Q.~M., \& Ji, H.~S. \ 2013, \aap, 557, L5
\bibitem[Zhou \& Ji(2009)]{Zhou09} Zhou, T.~H., \& Ji, H.~S.\ 2009, Research in Astronomy and Astrophysics, 9, 323
\bibitem[Zhou et al.(2013)]{Zhou13} Zhou, T.~H., \& Wang, J.~F., \& Li, D., et al.\ 2013, Research in Astronomy and Astrophysics, 13, 526
\end{thebibliography}
\end{document}